\documentclass{PoS}

\title{The lattice infrared Landau gauge gluon propagator: from 
finite volume to the infinite volume}

\ShortTitle{Lattice Infrared Landau Gauge Gluon Propagator}

\author{\speaker{O. Oliveira}$^1$, P. J. Silva$^{1,2}$ \\
        \llap{$^1$} Dep. F\'{i}sica, Universidade de Coimbra, 
3004-516 Coimbra, Portugal\\
        \llap{$^2$} School of Physics and Astronomy, University 
of Edinburgh, Edinburgh EH9 3JZ, UK\\
        E-mail: \email{orlando@teor.fis.uc.pt}, \email{psilva@teor.fis.uc.pt}}

\abstract{The Landau gauge lattice gluon propagator is discussed for different sets of lattices. Particular 
attention is given to its infrared properties. Our results show that the lattice propagator can be made 
compatible with either the decoupling-like or the scaling-like solution of the Dyson-Schwinger
equations. Furthermore, the analysis of the Cucchieri-Mendes bounds is performed considering
large volume simulations and the Oliveira-Silva ratios are computed. If the first do not give a clear answer about
the value of $D(0)$, the second method favors a $D(0) = 0$.
Finally, the SU(3) and SU(2) propagators are compared in the infrared. It comes out that the propagators
are different although the infrared exponents seem to be similar. 
The analysis suggests a scaling behaviour $D(0) \sim N$ with the gauge group SU(N).
}

\FullConference{International Workshop on QCD Green's Functions, 
Confinement, and Phenomenology - QCD-TNT09\\
		 September 07 - 11 2009\\
		 ECT Trento, Italy}

\usepackage{graphicx}

\begin{document}

%====================================================================================
%====================================================================================
\section{Introduction}

For a given quantum field theory the Green's functions encode the  dynamical 
information. For QCD, in particular, the computation of any Green's function, 
such as the gluon propagator $D(q^2)$, over the entire momentum spectrum cannot rely on 
perturbation theory. In what concerns the gluon propagator, in the past years, there has been a discussion 
about its infrared behaviour and its value at zero momentum in the Landau gauge. 
The recent effort on computing $D(0)$ comes from its relation with the Gribov-Zwanziger gluon confinement 
mechanism, which implies $D(0) = 0$, and with the possibility of dynamical mass generation for gluons. 
The dispute is still going on and involves both Schwinger-Dyson solutions and lattice QCD results. 

The recent solutions of the Dyson-Schwinger equations have different infrared 
behaviours. On one side we have the scaling solution \cite{Lerche,Fischer}
with $D(0) = 0$ and an infrared behaviour given by a 
pure power law $D(q^2) = (q^2)^{2\kappa -1}$, with $\kappa \sim 0.595$. On the other side we have the 
decoupling solution \cite{Aguilar,Papavassiliou} (see also \cite{Dudal})
with a finite and nonvanishing $D(0)$, with the value of $D(0)$  being related with a
dynamical generated gluon mass, and a plateau for $D(q^2)$ at low momenta.

The recent lattice simulations also show contradictory results. Indeed, large 
volume  simulations using the Wilson action and large lattice spacings, 
i.e. $a \sim 0.18$ fm or larger, show a gluon propagator that agrees 
qualitatively with the decoupling solution. However,  as we will see, 
a second look at the lattice data seems to indicate that the solution $D(0) = 0$ is still compatible with the lattice data --- see below.

Here, we want to discuss on what the lattice simulations tell us about the 
infrared gluon propagator and, hopefully, point towards a "favorite" value 
of $D(0)$. In the following, we will use standard definitions for each
of the quantities, which  will not be shown here. The interested reader can find the details
for example in \cite{SilvaOliveira04}.

%======================================================
%======================================================
\section{Lattice setup}

Our discussion of the infrared gluon propagator in the Landau gauge uses lattice data for the $SU(3)$ gluon
propagator from simulations with two different values of $\beta$, namely $\beta=6.0$ and $\beta=5.7$,
combining data generated at Coimbra with the data from the Berlin-Moscow-Adelaide group \cite{BMA}. 
Furthermore, the SU(3) data will be compared with the large volumes, i.e. $80^4$ and $128^4$ at
$\beta = 2.2$, SU(2) propagator of the S. Carlos group \cite{CuccLat07}.

\begin{figure}[h]
   \centering
   \includegraphics[scale=0.4]{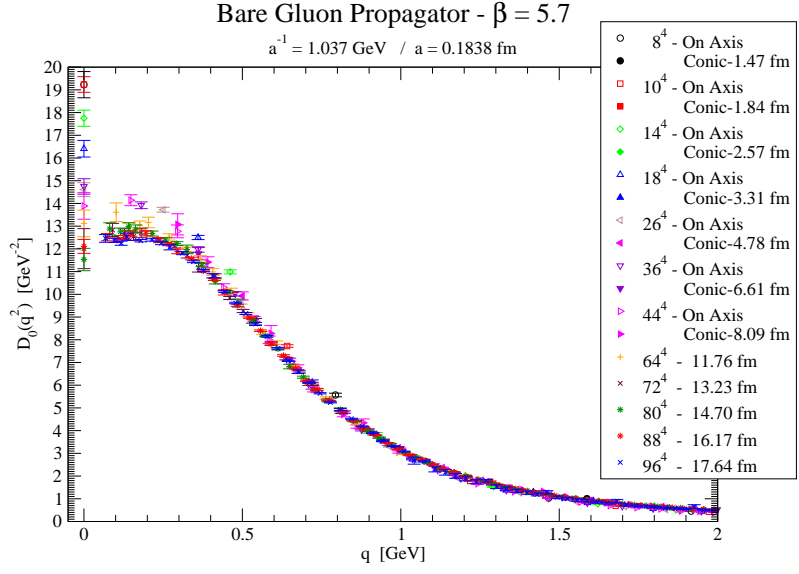} % requires the graphicx package
   \caption{Bare gluon propagator from $\beta = 5.7$ simulations. The data for the largest volumes at 
                    $L\geq64$ is the Berlin-Moscow-Adelaide data \cite{BMA} rescaled to our definition of the lattice
                    spacing.}
   \label{Db5.7}
\end{figure}

The lattices simulated using the SU(3) gauge group and the number of configurations for $\beta = 5.7$
are
\begin{table}[h]
   \centering
   \begin{tabular}{cccccccccccccc}     
    \hline
        L & 8 & 10	& 14 & 18 & 26 & 36 & 44 & 64$^*$ & 72$^*$ & 80$^*$ & 88$^*$ & 96$^*$ \\
       L(fm)     & 1.5 & 1.8 & 2.6 & 3.3 & 4.8 & 6.6 & 8.1 & 11.8 & 13.2 & 14.7 & 16.2 & 17.6 \\
       \# Confs &  56   & 149  & 149  & 149  & 132  & 100  & 29    & 14       & 20       & 25      &  68      & 67 \\
      \hline
   \end{tabular}
\end{table}

\noindent
where $^*$ stands for simulations carried out by the Berlin-Moscow-Adelaide group \cite{BMA}.
Note that we took the lattice spacing from the string tension \cite{Bali} and not from $r_0$ as the 
Berlin-Moscow-Adelaide group did in \cite{BMA}.  We have rescaled their data according to our definitions. 
We would like to call the reader's attention that different volumes have different statistics. 
The lattice gluon propagator for $\beta = 5.7$ can be seen in figure \ref{Db5.7}.

\begin{figure}[h]
   \centering
   \includegraphics[scale=0.4]{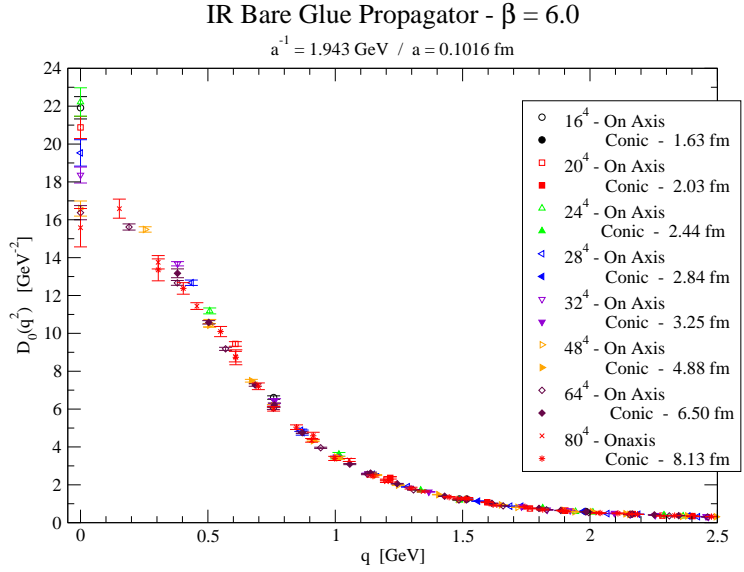} % requires the graphicx package
   \caption{Bare gluon propagator from $\beta = 6.0$ simulations.}
   \label{Db6.0}
\end{figure}

The simulations for $\beta = 6.0$, where the lattice spacing is $a=0.1016(25)$ fm,  were carried for the
following lattices 
\begin{table}[h]
\centering
\begin{tabular}{ccccccccc}
\hline
L   &  16 & 20 & 24  & 28 & 32  & 48 & 64 & 80 \\
L(fm)  &   1.6  & 2.0   & 2.4  & 2.8  & 3.2    &  4.9  & 6.5 & 8.1 \\
\# Confs &   52    & 72     &  60  & 56    & 126     &  104   & 120 & 18 \\
\hline
\end{tabular}
\end{table}

\noindent
and the propagators are plotted in figure \ref{Db6.0}.

As seen in figures \ref{Db5.7} and \ref{Db6.0}, no supression of the gluon propagator is observed.
Therefore, as a first and naive conclusion one could claim that, modulo finite volume effects,
the zero momentum gluon propagator is finite and non-zero.

%====================================================================================
%====================================================================================
\section{Modelling The Gluon Propagator \label{sec:fit}}

As a first step towards trying to distinguish between  a vanishing or nonvanishing $D(0)$, we look at the
compatibility of the lattice data with the functional forms which have been used to describe the two 
Dyson-Schwinger solutions, i.e. the infrared lattice data is fitted to
\begin{equation}
  D(q^2) ~ = ~ \frac{Z}{q^2 + M^2} \, ,
  \label{Ddecoupling}
\end{equation}  
where $M^2$ plays the role of a hard mass, and  
\begin{equation}
  D(q^2) ~ = ~ Z \, \frac{\left( q^2 \right)^{2 \kappa - 1}}{\left( q^2 + \Lambda^2 \right)^{2 \kappa} } \, .
  \label{Dscaling}
\end{equation}  
In the fits to (\ref{Dscaling}) the point $q = 0$ is not included. This is not unwise since finite volume effects are
certainly larger for $D(0)$.

\begin{figure}[htbp]
   \centering
  \includegraphics[scale=0.4]{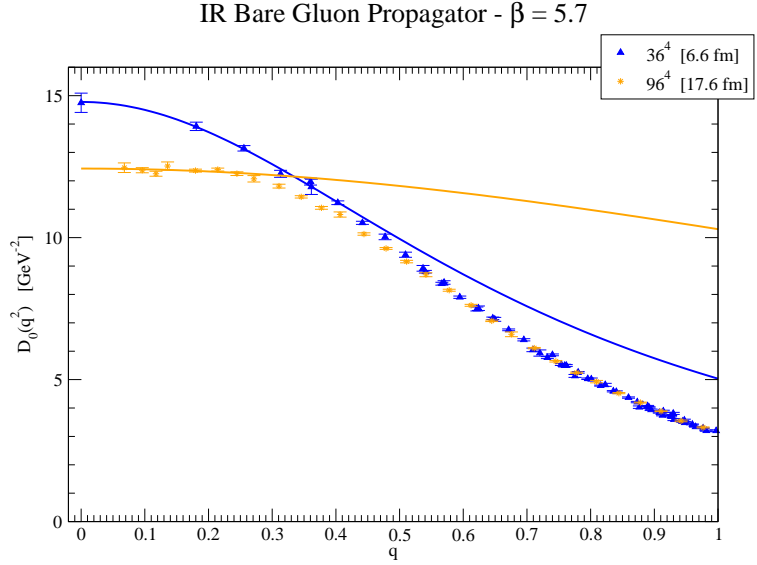} % requires the graphicx package
   \caption{IR bare gluon propagator for $\beta = 5.7$ and fits to (2.1).}
   \label{DMassiveFit}
   \end{figure}

\begin{figure}[htbp]
   \centering
   \includegraphics[scale=0.4]{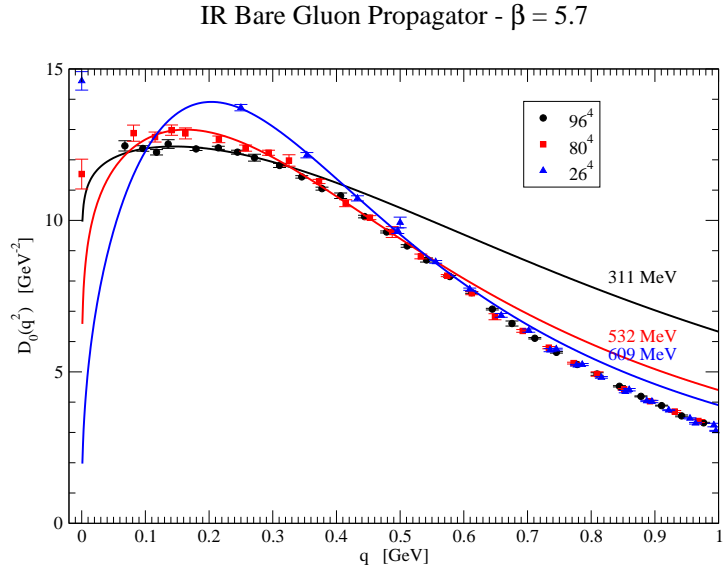} % requires the graphicx package
   \caption{IR bare gluon propagator for $\beta = 5.7$ and fits to (2.2). The MeV scales in the graph
   show the maximum fitting range compatible, i.e. with a $\chi^2/d.o.f. < 1.8$, with (2.2). Note that for the largest
   volume the $q = 0$ GeV point is missing.}
   \label{DScalingFit}
\end{figure}

In what concerns the fits to the $\beta = 5.7$ data, we have observed 
that for the massive like propagator
(\ref{Ddecoupling}), the smaller lattices ($L < 6 fm$) are not described 
by the above functional form. Moreover,
for the largest lattices, (\ref{Ddecoupling}) reproduces well the lattice 
data, i.e. fits have
$\chi^2/d.o.f. < 1.8$, for momenta up to 500 MeV with a gluon mass in the 
range of 719(18) MeV to
2.20(63) GeV. $M$ depends on the fitting range and on the lattice volume. 
It was observed that $M$ (and $Z$)
tend to increase with the lattice volume. For the fits to (\ref{Dscaling}), 
the results are similar. Indeed, the
lattice data is well described by (\ref{Dscaling}) to momenta up to 500 MeV, with 
$\Lambda$ decreasing from $\sim 900$ MeV to $\sim 420$ MeV as the lattice 
volume increases. The $\kappa$ is, within one standard deviation,
above 0.5 suggesting that $D(0) = 0$.
 Typical fits can be seen in figures \ref{DMassiveFit} and \ref{DScalingFit}.

\begin{figure}[htbp]
   \centering
   \includegraphics[scale=0.4]{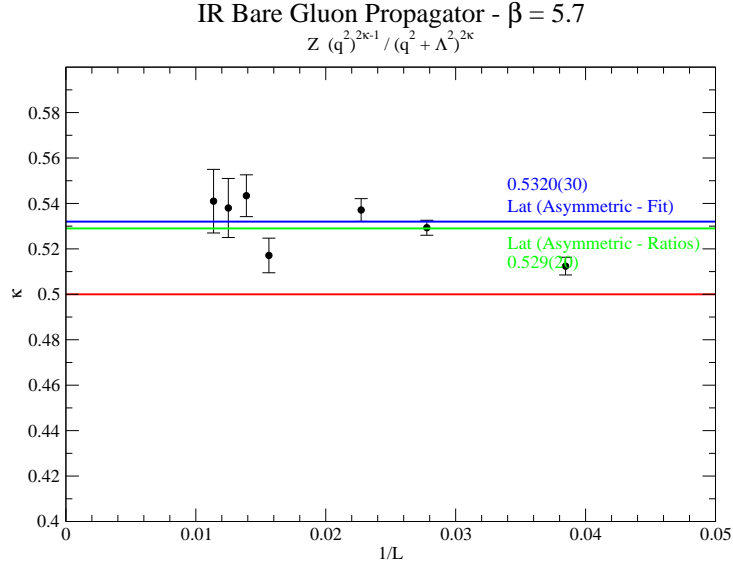} % requires the graphicx package
   \caption{Exponent $\kappa$ as measured from fitting (2.2) (black points). The plot also shows $\kappa$
   measured using $\beta = 6.0$  asymmetric lattice data to fit (2.2) (blue line) and  ratios of propagators 
   (green line) using the same asymmetric lattice set. The red line is $\kappa = 0.5$ and a value of $\kappa$
   above 0.5 implies $D(0) = 0$, a $\kappa < 0.5$ means $D(0) =  \infty$ and $\kappa = 0.5$ implies a 
   decoupling
   type solution.}
   \label{DscalingKappa}
\end{figure}

We repeated the procedure using the SU(2) gluon propagator of \cite{CuccLat07} and  found a similar behaviour. The main difference being that the SU(2) mass scales are typically larger than the corresponding 
SU(3) mass scales. For $M$ we get numbers around 1 GeV, while $\Lambda$ stays between 700 MeV to
800 MeV and $\kappa$ touch values slightly below the "magics" 0.5.

The results of fitting the $\beta = 6.0$ lattice gluon data give similar results. For completeness we resume the
results as follows:
\begin{enumerate}
\item "Decoupling" Fit: 

$48^4$ (L=4.9 fm), $q$ up to 508 MeV; $M$ goes from 600(20) MeV down to 578(12) MeV as the fitting range is increased;

$64^4$ (L=6.5 fm), $q$ up to 503 MeV; $M$ goes from 753(57) MeV down to  655.8(9.3) MeV as the fitting range is increased;

$80^4$ (L=8.1 fm), $q$ up to 664 MeV; $M$ goes from 588(129)  MeV down to  576.4(6.6) MeV as the fitting range is increased;

\item "Scaling" Fit:

$48^4$ (L=4.9 fm), $q$ up to 671 MeV; $\Lambda$ goes from 460(61)  MeV down to  432(27)  MeV and
$\kappa$ goes from 0.579(56) to 0.606(34) as the fitting range is increased;

$64^4$ (L=6.5 fm), $q$ up to 503 MeV; $\Lambda$ goes from 609(111)  MeV up to   614(43)  MeV and
$\kappa$ goes from 0.510(31) to 0.513(16) as the fitting range is increased;

$80^4$ (L=8.1 fm), $q$ up to 645 MeV; $\Lambda$ goes from 1.1(1.8) GeV  down to  582(54) MeV and
$\kappa$ goes from 0.453(61)  to  0.556(17) as the fitting range is increased
\footnote{ The reader should be aware that
the presented results for the $80^4$ lattice have a quite small statistics. So far, the total number of gauge configurations is 18, which could explain the relative large errors in $\Lambda$ and the rather low value for $\kappa$ obtained with the smallest fitting range.}.
\end{enumerate}

From the previous analysis, the conclusion is that although a "decoupling" like gluon propagator seems to
be favoured by the raw lattice data, a "scaling" like propagator is not excluded yet. The fits just described
show that both type of solutions are in good agreement with the lattice data and that
if a "scaling" like propagator is the solution, $D(q^2)$ starts to be supressed only for rather low momenta. Believing on the results of figure \ref{DScalingFit}, $D(q^2)$ is supressed only for momenta well below 
100 MeV.

%====================================================================================
%====================================================================================
\section{Cucchieri-Mendes Bounds \label{sec:cmb}}

In \cite{CMBounds}, the authors derived inequalities between $D(0)/V$ 
and what they called an average absolute value of the components of 
the colour magnetization $M(0)$,
\begin{equation}
 \langle M(0) \rangle^2 \le \frac{D(0)}{V} \le d \left( N^2_c - 1 \right) 
\langle M(0)^2 \rangle \, ,
 \label{CMB}
\end{equation}
where $d$ is the number of space-time dimensions and $N_c$ the number of colours.
For the definition of $M(0)$ see the cited work. 
In the above expression, $\langle \cdots \rangle$ means Monte Carlo average 
and  (\ref{CMB}) follows directly from using a Monte Carlo approach. 

In \cite{CMBounds} the different 
terms in
(\ref{CMB}) were computed for SU(2) simulations 
and, after performing a scaling analysis, the authors conclude 
in favour of a finite and nonvanishing $D(0)$.
Indeed, they claimed $D(0) \ge 2.2(3)$ GeV$^{-2}$. A similar analysis 
was performed for SU(3) in 
\cite{CMBSU3} and the conclusions favour a $D(0) = 0$, 
although a finite nonvanishing $D(0)$ was not 
completely excluded. The simulations use the Wilson action 
and different lattice spacings and volumes. For the SU(2) simulations 
the authors used $a \sim 0.22$ fm and volumes up to (27 fm)$^4$, 
whereas 
the SU(3) simulations were performed with $a \sim 0.10$ fm and volumes 
up to (6.5 fm)$^4$.

Using the lattice data described before for $\beta = 5.7$, 
we are now in position of review the scaling analysis for the 
$SU(3)$ gauge theory --- see also \cite{Lat09}. Note that for the largest 
volumes $64^4$ - $96^4$, which are data from the Berlin-Moscow-Adelaide group, 
we only have access to $D(0)$.  

\begin{figure}[t]
   \centering
   \vspace*{0.4cm}
   \includegraphics[scale=0.4]{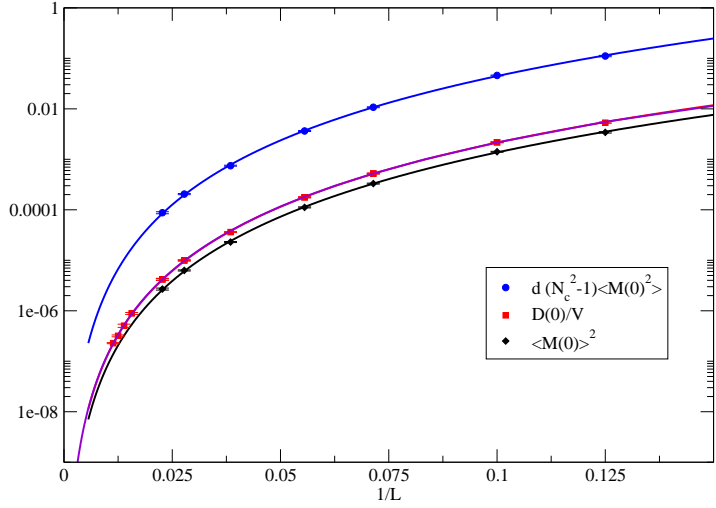} % requires the graphicx package
   \caption{Cucchieri-Mendes bounds, the lattice data and corresponding fits to $A/V^\alpha$. Note that in all fits the $26^4$ data were excluded in order to have a $\chi^2/d.o.f. < 1.8$. Furthermore, for $D(0)/V$ the plot includes the fits  to the smaller set of lattices, i.e. $8^4 - 44^4$ and to all data available, i.e. to the lattices $8^4 - 88^4$. The agreement between the two sets of data is perfect.}
   \label{Bounds57Fits}
\end{figure}

Anyway, following \cite{CMBounds,CMBSU3}, we assume that in (\ref{CMB}) the 
different functions depend on the lattice volume as $A/V^\alpha$. Then, 
it follows that an $\alpha > 1$ means $D(0) = 0$ in the
infinite volume. The fits of the lattice data to the small set of volumes 
($L \le 8.1$ fm) give
\begin{table}[h]
   \centering
   \begin{tabular}{lccc}   
    \hline
                        &  $\langle M(0) \rangle^2 $ &  $D(0)/V$       & $\langle M(0)^2 \rangle$ \\
     \hline
   $\alpha$     &              1.0537(50)              &   1.0504(45)  &  1.0530(50) \\
      \hline
   \end{tabular}
\end{table}

\noindent
and confirm the 
conclusions presented in \cite{CMBSU3}, as we found $\alpha>1$.
If one wants to use the full set of $\beta = 5.7$ lattices, we can only
investigate the scaling behaviour of $D(0)/V$. In this case, the fit 
gives $\alpha = 1.0538(28)$ with a $\chi^2/d.o.f. = 0.87$,
which is in excelent agreement with the estimation using the 
smaller set of lattices, and again it suggests a 
$D(0) = 0$ in the infinite volume. The $\beta = 5.7$ lattice data 
and the corresponding fits to $A/V^\alpha$
can be seen in figure \ref{Bounds57Fits}.
Again, as in  \cite{CMBSU3}, if one assumes that the dependence with the
volume of the different members of  (\ref{CMB}) are given by $C/V + D/V^\alpha$, 
then the data is well described by this functional form;
in this sense, a finite and non-vanishing value for $D(0)$
in the infinite volume is not excluded.

%====================================================================================
%====================================================================================
\section{Power-law Behaviours From Ratios Of Propagators}

In \cite{EPJC} a method was proposed which tests the compatibility of the lattice data with a power law
behaviour and, simultaneously, suppresses the finite volume effects. Since the method is not wellknown, for completeness, we will review it. The infrared propagator can be investigated using on-axis momenta which
are defined as
\begin{equation}
   q[n]  =  \frac{2}{a} \sin\left( \frac{\pi n}{L} \right), \hspace{1cm} n = 0, 1, \dots, \frac{L}{2}
\end{equation}
for a symmetry $L^4$ lattice. If the gluon propagator is described by a pure power law, i.e.
\begin{equation}
D(q^2) ~ = ~ Z \, \left( q^2 \right)^{2 \kappa - 1} \, ,
\end{equation}
one can define the following ratio
\begin{eqnarray}
  R[n] & = & \ln \left[ \frac{ q^2[n+1] \, D( q^2[n+1]  ) }{q^2[n] \, D( q^2[n]  )} \right] \\
          & = & 2 \kappa \, R_q [n] ~ = ~ 2 \kappa \,  \ln \left[ \frac{ q^2[n+1] }{q^2[n] } \right]  \, .
\end{eqnarray}

\begin{figure}[b]
   \centering
   \vspace*{0.4cm}
   \includegraphics[scale=0.4]{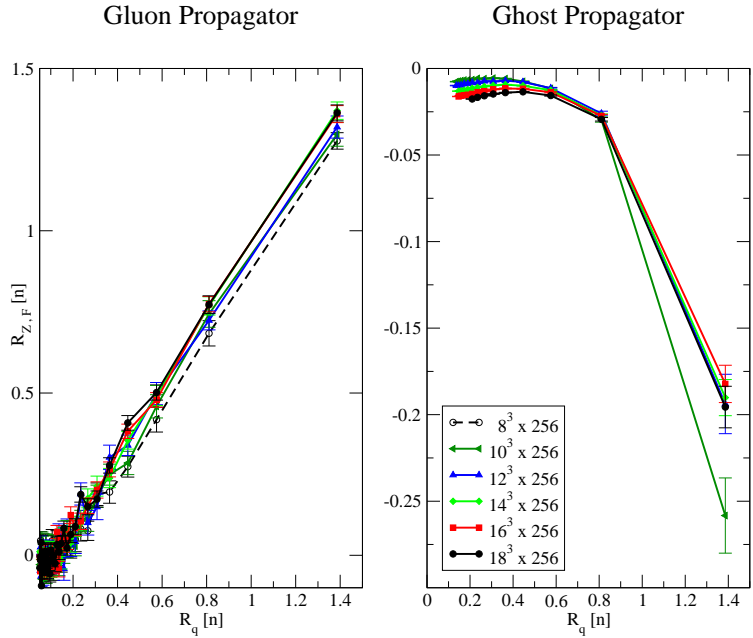} % requires the graphicx package
   \caption{Ratios of gluon and ghost propagators for large asymmetric lattices \cite{EPJC}. Note that, while the gluon data
   seems to be compatible with a power law behaviour, the ghost propagator does not 
seem to
follow a power law.}
   \label{RatiosAsy}
\end{figure}

The compatibility of the lattice data with a power law can be tested fitting the lattice data to
\begin{equation}
  R[n] ~ = ~ 2 \kappa \, R_q [n] \, + \, C \, ,
  \label{Rr}
\end{equation}
where $C$ is a constant which resumes both the deviations 
from a power law behaviour
and the finite volume effects
which are not eliminated by taking the ratio of propagators in $R[n]$. In \cite{EPJC}, the authors analyzed such 
type of fits to the lattice data for large asymmetric $8^3 - 18^3 \times 256$ lattices --- see figure \ref{RatiosAsy}.
For the gluon propagator the fits give, within one standard deviation, $\kappa >  0.5$  with $\kappa \sim 0.53$
and a constant $C$ which seems to 
approach zero
as we go from $8^3 \times 256$
to $18^3 \times 256$. The first result suggests a $D(0) = 0$, while the second result suggests that $C$ 
mainly resumes
the finite volume effects. In what concerns the ghost propagator, 
figure \ref{RatiosAsy} shows that either the data is still far from the linear behaviour or the ghost propagator does not follow a pure power law in the infrared region.

\begin{figure}[t] 
   \centering
   \includegraphics[scale=0.4]{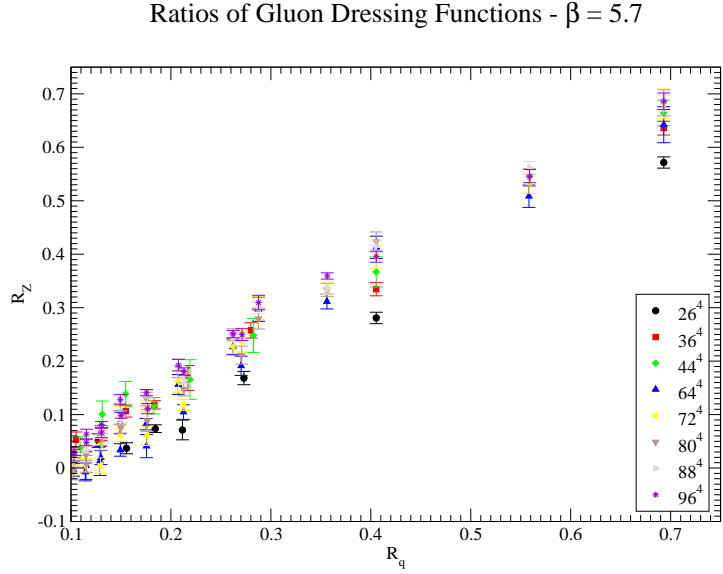} 
   \caption{Ratios of propagators for $\beta = 5.7$ lattices.}
   \label{RB57}
\end{figure}

\begin{figure}[t] %  figure placement: here, top, bottom, or page
   \centering
   \includegraphics[scale=0.4]{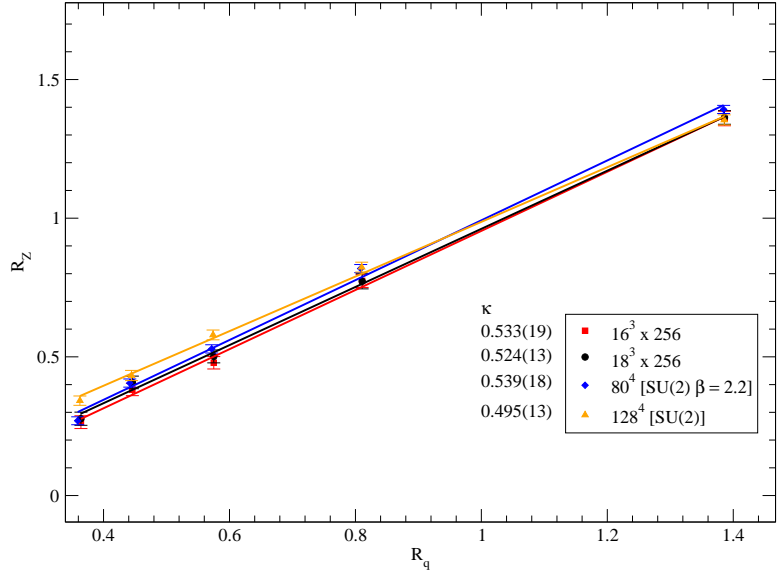} 
   \caption{Ratios of propagators for SU(3) asymmetric lattices at $\beta=6.0$ and SU(2) simulations.}
   \label{Rsu2}
\end{figure}

The ratios for the $\beta = 5.7$ propagators can be seen in figure \ref{RB57}. 
We have tried to fit the
data coming from the largest lattice volumes 
to  (\ref{Rr}) but it turns out that the $\chi^2/d.o.f.$ were always too large,
 i.e. well above 2. This is probably
due to rotational invariance violations as the infrared 
Berlin-Moscow-Adelaide data mixes different types 
of momenta which have different types of corrections due to 
the lack of rotational invariance.
Indeed  a similar effect is seen in results obtained from other 
lattices when one mixes on-axis with off-axis momenta.
Given that we cannot distinguish between the two types of points in 
the Berlin-Moscow-Adelaide data, we show in figure \ref{Rsu2} the ratios 
for some asymmetric lattices and the large volume SU(2) 
propagators. The figure includes the $\kappa$ measured from fitting 
the lattice data to (\ref{Rr}). 
The $128^4$ results for the SU(2) gauge group is the only data which 
gives a $\kappa < 0.5$, within one standard 
deviation. However, the infrared $128^4$ propagator, see figure 
\ref{Rsu2su3Prop}, shows larger fluctuations than
any other calculation. Furthermore, it is one of the very few 
simulations where one sees an 
enhancing of the 
propagator in infrared region. Certainly, this is due to the small statistics.

%====================================================================================
%====================================================================================
\section{How Different are the SU(2) and SU(3) Infrared Gluon Propagators?}

As a final topic, we would like to discuss how different are the SU(2) and 
SU(3) propagators. The discussion of section \ref{sec:fit} suggests that there should be
a difference in the infrared. Indeed, the mass scales for SU(2) and SU(3)
don't seem to be equal with $M_{SU(2)} > M_{SU(3)}$. 
Moreover, the scaling analysis of the Cucchieri-Mendes bounds seems to give 
different conclusions when we use different gauge groups --- see section \ref{sec:cmb} and \cite{CMBounds, CMBSU3}.
However, in \cite{su2su3,australianos} it was shown that the two 
propagators are similar for momenta larger than $\sim$ 800 MeV. 
For smaller momenta, one can see some differences whose origin 
was not clear. If one uses now the SU(2) data from the S. Carlos 
group and the 
SU(3)
Berlin-Moscow-Adelaide data one can compare results 
for volumes up to (17 fm)$^4$. In order to compare
the SU(2) and SU(3) propagators, they were renormalized by the condition
\begin{equation}
   \left.  D(q^2) \right|_{q^2 = \mu^2} ~ = ~ \frac{1}{\mu^2}
   \label{renor}
\end{equation}
choosing $\mu = 3$ GeV. The renormalization constants were 
computed after fitting $D(q^2)$ to the 1-loop
perturbative result for $q > 2.5$ GeV. In these fits for SU(3) 
we used the conic cut data, while for the SU(2) we used only the 
diagonal momenta. The renormalization constants were computed from the 
condition (\ref{renor}) using the results of the fits. The statistical 
error in the renormalization constants being around 10\%. 

\begin{figure}[t] %  figure placement: here, top, bottom, or page
   \centering
   \includegraphics[scale=0.4]{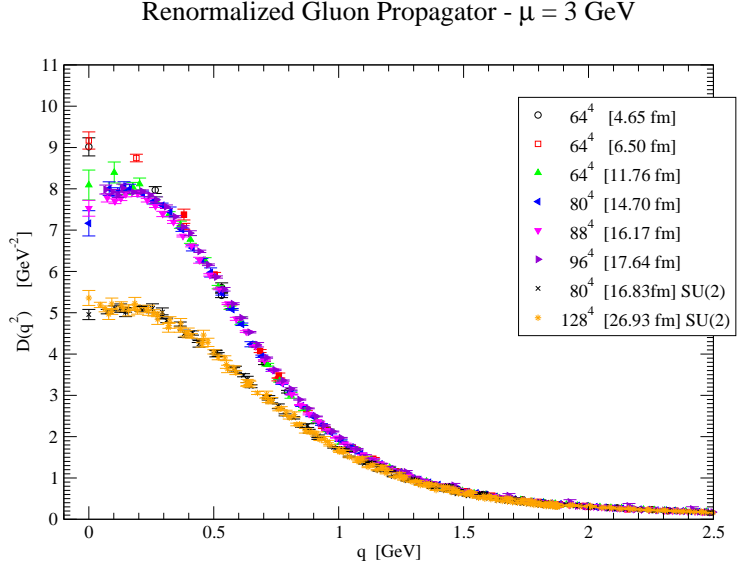} 
   \caption{Comparing SU(2) and SU(3) gluon propagators.}
   \label{Rsu2su3Prop}
\end{figure}   

The renormalized propagators, which 
can be seen in figure \ref{Rsu2su3Prop}, 
show clear differences in the infrared region. This result is not in contradiction with \cite{su2su3}, which had a limited access to the low momenta
region. Furthermore, reading $D(0)$ from figure \ref{Rsu2su3Prop} it follows that
\begin{equation}
   \frac{ D(0)_{SU(2)} } { D(0)_{SU(3)} } \sim \frac{5}{7.5} = \frac{2}{3} \, ,
\end{equation}
suggesting a $D(0)_{SU(N_c)} \sim N_c$. This result is not necessarily in conflict with the large $N_c$ 
expansion because the gluon propagator is not a gauge invariant object.

We would like to call the reader's attention that, although the propagators are different, the infrared exponents computed
by the ratio method discussed in the previous section seem to be similar.

%====================================================================================
%====================================================================================
\section{Conclusions}

The results discussed in this article show that 
results obtained from lattice simulations for the gluon propagator in Landau gauge
can be made
compatible with both types of Schwinger-Dyson solutions, i.e. it can be seen either as a scaling like solution,
where $D(0) = 0$, or as a decoupling solution, where $D(0)$ is finite and non-vanishing. The raw
lattice data shows a plateau at small momenta and, in this sense, 
it does look more like as a decoupling
type of propagator.

The analysis of the bounds derived in \cite{CMBounds} are not conclusive when applied to the SU(3)
simulations \cite{CMBSU3, Lat09}. According to the analysis shown here for the $\beta = 5.7$, the leading behaviour
is already captured when one includes volumes as "small" as $\sim$ (8 fm)$^4$. Moreover, the 
performed scaling analysis is
in good agreement with the ratios analysis, favouring a $D(0) = 0$. This does
not 
necessarily mean
that the lattice simulations point towards a scaling like solution. 
Remember that the ratio method applied to the asymmetric lattice data shows that the ghost propagator does
not follow a power law. In what concerns the gluon and ghost propagators, it looks like that before having
a clear understanding of the 
finite lattice volume/spacing
effects, it will be quite difficult to give a definitive answer on the
nature of the propagators in the deep infrared region.

Finally, 
the Landau gauge gluon SU(2) and SU(3) propagators are compared for momenta below 800 MeV. It turns
out that there are clear differences and a scaling law $D(0)_{SU(N_c)} \sim N_c$ seems to hold.

%====================================================================================
%====================================================================================
\section*{Acknownledgements}

We would like to thank the Berlin-Moscow-Adelaide and S. Carlos groups for sending us their data and for allowing to use it.

The authors also acknowledge financial support from FCT under projects CERN/FP/83582/2008 and CERN/FP/83664/2008. 
P. J. S. was supported by FCT via grant SFRH/BPD/40998/2007. 

%====================================================================================
%====================================================================================

\end{document}